\documentclass[reprint,prl,notitlepage,twocolumn,
superscriptaddress,aps,longbibliography,floatfix]{revtex4-1}

\usepackage{braket,color}
\usepackage{amsmath}
\usepackage{amsfonts}
\usepackage{dsfont}
\usepackage{bm}
\usepackage[utf8]{inputenc}
\DeclareMathOperator{\tr}{tr}
\usepackage{amsthm}
\newtheorem{thm}{Theorem}[]
\newtheorem{cor}{Corollary}[]
\newtheorem{proposition}[thm]{Proposition}
\newtheorem{lemma}[thm]{Lemma}
\usepackage{graphicx}

\begin{document}

\title{Efficient Learning of Continuous-Variable Quantum States}
\author{Ya-Dong Wu}
\affiliation{John Hopcroft Center for Computer Science, Shanghai Jiao Tong University, Shanghai 200240, China}
\affiliation{QICI Quantum Information and Computation Initiative, Department of Computer Science,
The University of Hong Kong, Pokfulam Road, Hong Kong}
\author{Yan Zhu}
\affiliation{QICI Quantum Information and Computation Initiative, Department of Computer Science,
The University of Hong Kong, Pokfulam Road, Hong Kong}
\author{Giulio Chiribella}
\affiliation{QICI Quantum Information and Computation Initiative, Department of Computer Science,
The University of Hong Kong, Pokfulam Road, Hong Kong}
\affiliation{Department of Computer Science, Parks Road, Oxford, OX1 3QD, UK}
\affiliation{Perimeter Institute for Theoretical Physics, Waterloo, Ontario N2L 2Y5, Canada}
\author{Nana Liu}
\affiliation{Institute of Natural Sciences, Shanghai Jiao Tong University, Shanghai 200240, China}
\affiliation{Ministry of Education, Key Laboratory in Scientific and Engineering Computing, Shanghai Jiao Tong University, Shanghai 200240, China}
\affiliation{School of Mathematical Sciences, Shanghai Jiao Tong University, Shanghai, 200240, China}
\affiliation{University of Michigan-Shanghai Jiao Tong University Joint Institute, Shanghai 200240, China}

\begin{abstract}
        The characterization of continuous-variable quantum states is crucial for applications in quantum communication, sensing,  simulation and  computing. However, a full characterization  of multimode quantum states requires a number of experiments  that  grows  exponentially with the number of modes. Here we propose an alternative approach where the goal is  not to reconstruct the full quantum state, but rather to estimate its characteristic function at a given set of points. %For  states  with reflection symmetry, such as Fock states, Gottesman-Kitaev-Preskill (GKP) states,  Schrödinger cat states, and Gaussian states with zero mean values,  our strategy 
        %only requires homodyne measurements and balanced beam splitters performed on a constant number of copies of the given state. 
       For multimode states with reflection symmetry, we show that the characteristic function  at $M$ points  can be estimated  using only  $O(\log M)$ copies  of the state, independently of the number of modes. When the characteristic function is known to be positive, as in the case of squeezed vacuum states, the estimation is achieved by an experimentally friendly setup using only beamsplitters and homodyne measurements. 
\end{abstract}

\maketitle

{\em Introduction.}
%Recently, remarkable quantum advantage has been shown in learning quantum states and quantum circuits~\cite{huang2021,huang2022}, indicating that detecting quantum properties through coherent access of quantum copies is much more efficient than classical information processing following quantum measurements on each single quantum copy.
%The results of quantum advantage in learning quantum systems are established on many-qubit systems, however, it is unknown whether quantum advantage also exists in learning CV quantum states.
Continuous-variable (CV) quantum systems~\cite{RevModPhys.77.513,RevModPhys.84.621} are an important platform for quantum computing, simulation, sensing, and communication. 
In quantum computing, CV systems play  a significant role in quantum error correction and fault-tolerance~\cite{gottesman2001,mirrahimi2014,michael2016}, and  have been shown to offer  quantum  advantages in sampling problems~\cite{hamilton2017,douce2017,madsen2022}.
  They also feature in  quantum machine learning, where they provide a platform for realization of quantum neural networks~\cite{killoran2019,arrazola2019}. 

A large body of work has been devoted to the characterization of CV quantum states, exploring a variety of techniques including quantum  tomography~\cite{ariano2003,lvovsky2009}, quantum compressed sensing~\cite{ohliger2011}, quantum fidelity estimation~\cite{silva2011}, detection of nonclassicality~\cite{mari2011} and certification of quantum states~\cite{Aol15,wu2021,chabaud2021,wu2022certification}.  Classical shadow tomography has also been generalized to CV states~\cite{iosue2024,gandhari2024,becker2024}, however they require truncation of infinite-dimensional Hilbert space and the analysis of their computational complexity is absent. Recently,  classical machine learning techniques  have been applied to the characterization of CV states~\cite{tiunov2020,cimini2020,Ahmed2021PRL,Ahmed2021PRR,hsieh2022,zhu2022,wu2023}.  These  approaches have been shown to achieve high quality performance on specific CV states,  although in general it is hard to provide  rigorous {\em a priori} guarantees on their error scaling and sample complexity.  
%In all these approaches, individual quantum measurements are performed on each copy of an unknown quantum state.  A natural question is whether we can achieve an advantage by performing global measurements on multiple copies of state. 

%In conventional quantum state tomography, given copies of unknown state $\rho$, the aim is to fully estimate its density matrix. 

The full characterization of a multimode quantum state generally requires measurements on  an exponential number of  copies of the state, and therefore becomes unfeasible when the number of modes is large.  Here, we explore an alternative approach, where the goal is not to completely characterize the state, but rather to estimate its characteristic function at a finite number of points. The characteristic function is also important in the study  of quantum information scrambling in phase space~\cite{zhuang2019}, and its estimation is often used as the first step in experimental schemes of CV  state tomography~\cite{PhysRevLett.125.043602,campagne2020quantum,eickbusch2022}.   Furthermore, estimates of  the characteristic function can also be used  for learning  nonlinear and global  properties of multimode quantum states,  such as   amount of nonclassicality~\cite{mari2011}, non-Gaussianity~\cite{cimini2020,mattia2021},  or the fidelity with a given target state~\cite{silva2011}.  
%[is there something fancier  than the purity? After all, purity can be estimated "easily" with a control swap.]

In this paper, we develop an efficient method for  estimating point values of the characteristic function of a multimode quantum state.
For CV states with reflection symmetry, we show that the square of the characteristic function can be estimated with homodyne measurements and balanced beamsplitters using a finite number of copies of the state.   Building on this result, we then show that the characteristic function at $M$ points can be estimated using only $O(\log M)$ copies of the state, using suitable global measurements.  Notably, the sample complexity is independent of the number of modes, and unlike in classical shadow tomography,   our method does not require  a truncation of the Hilbert space, neither in the Fock basis nor in phase space.  Furthermore, it turns out that our approach has a lower  computational complexity compared to traditional  quantum state tomography or its more recent versions,  and that the complexity scales as $O(M \log M)$ with the number of points.

{\em Background.}  To get around the exponential complexity of quantum state tomography,  Aaronson proposed the method of shadow tomography, which predicts  the expectation values of a  set of observables~\cite{aaronson2018}.
Building on this result, Huang {\em et al.}\ proposed classical shadow tomography~\cite{huang2020},
 which has been recently extended to CV quantum states~\cite{iosue2024,gandhari2024,becker2024}. 
When used to estimate  the expectation values of any of all $4^n$ Pauli observables on an $n$-qubit state, however, classical shadow tomography still requires an exponential number of measurements. To provide an efficient estimate of all Pauli observables, a quantum strategy using global measurements on multiple copies was then shown ~\cite{huang2021}. In the following we will establish an analogue result for CV systems, with the crucial difference that instead of estimating the expectation values of an arbitrary set  of observables, we will estimate the values of the characteristic function at an arbitrary set of phase space points. 
%{\color{blue} [actually, I am not sure we need this figure.]} 

Consider a $k$-mode quantum system, described by the Hilbert space $\bigotimes_{j=1}^k\mathcal H_j$ where each $\mathcal H_j$ is an infinite-dimensional Hilbert space, associated with the annihilation and creation operators $a_j$ and $a_j^\dag$, respectively.    A multimode displacement operator  is a unitary operator of the form $D(\bm{\alpha})=\textrm{e}^{\bm{\alpha} \hat{\bm{a}}^\dagger-\bar{\bm{\alpha}} \hat{\bm{a}}}$, where $\bm{\alpha}=(\alpha_1, \dots, \alpha_k)\in \mathbb C^k$, $\hat{\bm{a}}=(\hat{a}_1, \dots, \hat{a}_k)^\top$, $\hat{\bm{a}}^\dagger=(\hat{a}_1^\dagger, \dots, \hat{a}_k^\dagger)^\top$, and the annihilation and creation operators satisfy  the  canonical communication relations $\left[\hat{a}_j, \hat{a}^\dagger_l\right]=\delta_{jl}$ for every $j$ and $l$.  

The characteristic function of a quantum state $\rho$ is defined as $C_\rho(\bm{\alpha}):=\tr[D(-\textrm{i}\bm{\alpha})\rho]$~\cite{leonhardt1997}.   It fully characterizes the quantum state $\rho$, which can be reconstructed with  the tomographic formula  $\rho =1/\pi^k\int_{\mathbb C^k} d^{2k}\bm{\alpha} \,  C_\rho  (\bm {\alpha})  \, D(i\bm{\alpha})$.  In spin-boson systems, the characteristic function of a bosonic mode at each phase space point can be directly sampled by using a state-dependent displacement operation followed by measuring the spin~\cite{PhysRevLett.125.043602}. The characteristic function can also be obtained from the Wigner function, which is often used to represent CV  states and can  be reconstructed from its marginals by homodyne measurements in optical systems~\cite{lvovsky2009},  via an inverse Fourier transform in phase space~\cite{scully1997,leonhardt1997,serafini2017}.  

A simple way to estimate the characteristic function at a specific point $\bm \alpha$ is to subject   each mode $j$ to a homodyne measurement, {\em i.e.} a  projective measurement of the quadrature operator $Q_{\alpha_j}   := ({\alpha_j}  \, \hat{a}_j^\dag + \overline{{\alpha}}_j  \, \hat{a}_j)/|\sqrt 2 \alpha_j|$.  From the value of the measurement outcome $q_j$,  one can then evaluate the empirical average of the exponential $\exp[-i \sum_{j=1}^kq_j  \,  |\sqrt 2 \alpha_j|]$, which provides an estimate of the characteristic function when averaged over many repetitions of the measurement procedure.  
% This approach is commonly used in quantum state tomography due to its ease of implementation~\cite{lvovsky2009}. 
However, this approach has the obvious limitation that the sample complexity grows  linearly with the number of points  where the characteristic function is evaluated.     In the following, we provide an exponentially more efficient method.
 %working on a class of physically relevant CV states.

\medskip

 {\em Efficient estimation of the characteristic function.}    Our method applies to quantum states with {\em reflection symmetry}, that is, quantum states $\rho$ for which there exists a  $k\times k$ unitary matrix $U$ such that $C_\rho  (-\overline{\bm \alpha})   =  C_\rho  (  {\bm \alpha}   U)$  for every vector of displacements $\bm \alpha$.    In the single-mode case,  quantum states with reflection symmetry  include important classes of states such as Gaussian states with zero mean values, Fock states, Gottesman-Kitaev-Preskill (GKP) states~\cite{gottesman2001}, Schrödinger cat states~\cite{mirrahimi2014} and  binomial code states~\cite{michael2016}.

 Our main result is the following theorem, which provides a method for  estimating the characteristic function of a multimode state with reflection symmetry.  The sample complexity of our estimation strategy is independent of the number of modes, and logarithmic in the number of evaluation points. 
 
\begin{thm}\label{thm}
    For every $k$-mode state $\rho$ with reflection symmetry,   the values  of the characteristic function  $C_\rho  (\bm{\alpha} )$  at  $M$ given points $\{\bm{\alpha}_i\}_{i=1}^M$ can be accurately estimated  using  $O(\log M)$ copies, independently of $k$.  Specifically, $O(1/\epsilon^4\log(M/\delta))$ copies are sufficient to produce an  estimate  $\widehat{C_\rho({\bm{\alpha}})}$   that satisfies the condition  
     ${\sf Prob} \left(\max_i  \, \left| \widehat{C_\rho({\bm{\alpha}_i})}- C_\rho(\bm{\alpha}_i)\right|>\epsilon\right)<\delta$ for any fixed $\epsilon>0$ and $\delta > 0$. 
\end{thm}

Here $\delta$ denotes the failure probability that at least one estimation is $\epsilon$-far from the true value. The theorem is based on two techniques, which are interesting in their own right. The first  technique allows one to  estimate the product $C_\rho  (\bm{\alpha})  C_\rho  (-\overline{\bm{\alpha}})$  for an arbitrary CV state $\rho$, without any assumption of reflection symmetry.  The measurements  used in the estimation  are experimentally friendly,  requiring only beamsplitters and homodyne detections. We note that our estimates are for the ideal experimental setup and we leave more realistic estimates with photon losses and other experimental efficiencies to future work.\\

The sample complexity of this strategy is constant in the number of modes, and depends only on the chosen error threshold:

\begin{lemma}\label{lemma1}
 For every $k$-mode state $\rho$, $O(\log(1/\delta)/\epsilon^2)$  copies    of $\rho$  are sufficient  to produce  an estimate $\widehat{C_\rho({\bm{\alpha}})C_\rho(-\bar{\bm{\alpha}})}$ that satisfies the condition   ${\sf Prob}\left(  \left| \widehat{C_\rho({\bm{\alpha}})C_\rho(-\bar{\bm{\alpha}})}- C_\rho(\bm{\alpha}) C_\rho(-\bar{\bm{\alpha}})\right|>\epsilon\right)<\delta \, ,\forall \alpha \in \mathbb C^k$.  The protocol and its sample complexity are independent of $\bm \alpha$.
\end{lemma}

\begin{figure}
    \centering
    \includegraphics[width=0.45\textwidth]{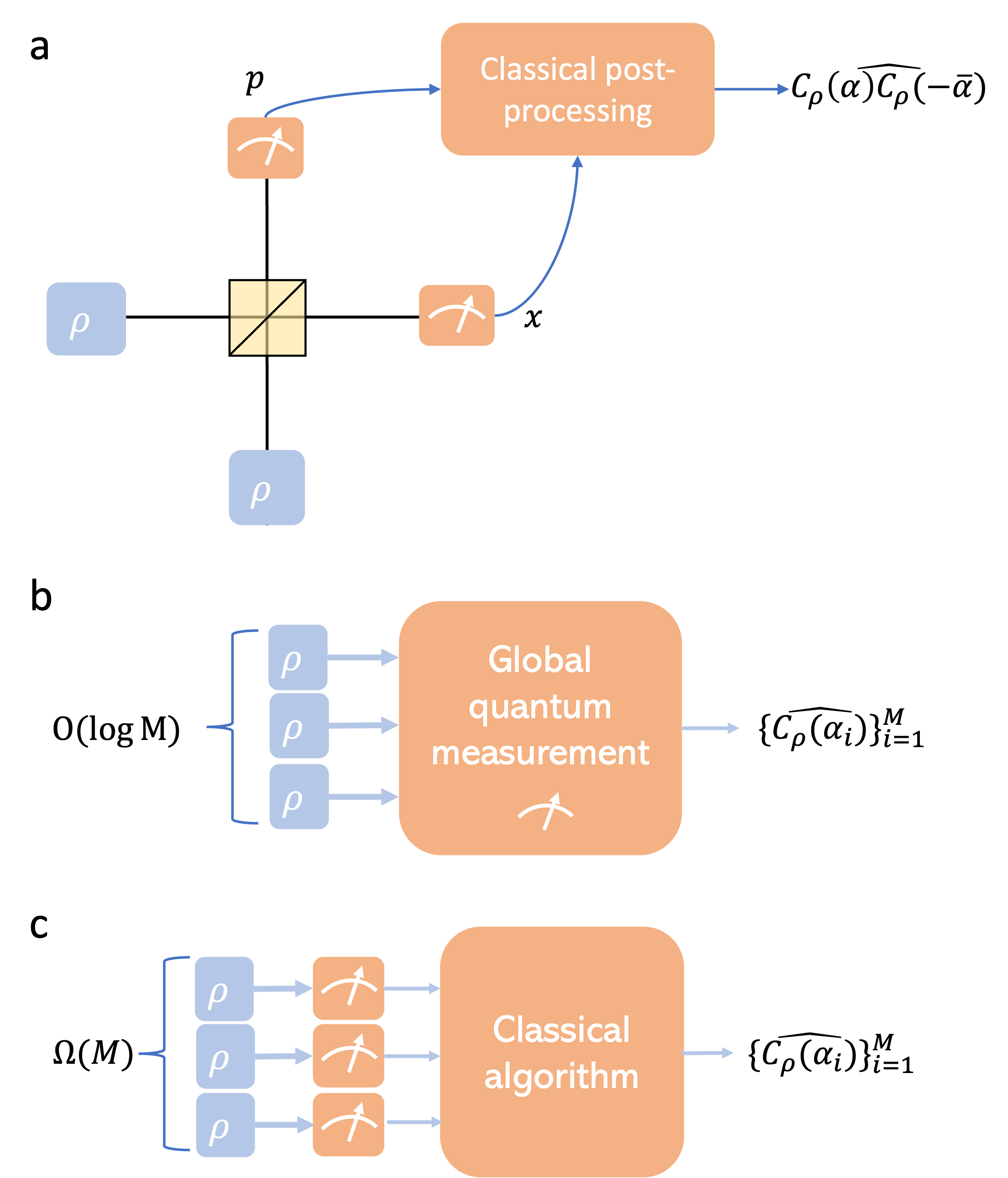}
    \caption{ Schemes for estimating the characteristic function.    Subfigure a shows our protocol for the estimation of $C_\rho  ( \alpha) C_\rho(-\overline{ \alpha})$  using a balanced beam splitter and two homodyne measurements. Subfigures b and c correspond to the  estimation of the characteristic function at $M$  phase-space points using global measurements, and conventional scenarios using single-copy measurements, respectively. %{\color{blue} maybe we can also put a $O(\log M)$ and $\Omega(M)$ before to the copies in subfigures b and c, so that one can see our method requires only $\log M$ copies, while the conventional one requires $M$. } 
    }
    \label{fig:my_label}
\end{figure}

 The idea  is that the product displacements  $D(-\textrm{i}\bm{\alpha})  \otimes D(\textrm{i}\overline{\bm{\alpha}})$ commute for all possible values of   $\bm{\alpha}$, and therefore are jointly measurable on the product state $\rho\otimes \rho$ at once.   In the single-mode case, the joint measurement is achieved by a simple setup, illustrated in Fig.~\ref{fig:my_label}a:  The product state $\rho \otimes \rho$ goes through  a balanced beam splitter followed by two homodyne detections on the two ouput modes, measuring on the spectral resolutions of the the position  operator $\hat{x}:=(\hat{a}+\hat{a}^\dagger)/\sqrt{2}$ and the momentum operator $\hat{p}:=(\hat{a}-\hat{a}^\dagger)/(\sqrt{2}\textrm{i})$, respectively.  Denoting the two measurement outcomes by $x$ and $p$ respectively, we have
\begin{equation}\label{eq:empiricalAvg}
     \braket{D(-\textrm{i}\alpha)\otimes D(\text{i}\bar{\alpha})}_{\rho\otimes \rho}
    %=& \left\langle D\left(-\textrm{i}\sqrt{2}\textrm{Re}(\alpha)\right)\otimes D\left(\sqrt{2}\textrm{Im}(\alpha)\right)\right\rangle_{\textrm{BS} \rho\otimes\rho \textrm{BS}^\dagger}\\
    =\mathds{E} \left[\textrm{e}^{-2\textrm{i} (\textrm{Re}(\alpha)x+ \textrm{Im}(\alpha)p)} \right]\, , 
\end{equation}
where %BS$=\textrm{e}^{\pi/4(\hat{a}_1^\dagger\hat{a}_2-\hat{a}_1\hat{a}_2^\dagger)}$ is the unitary operation of a balanced beam splitter, and
$\mathbb E$ denotes the expectation value over all possible pairs $(x,p)$ of measurement outcomes obtained in the experiment.
Using Hoeffding's inequality, one can then show that a constant number of copies of $\rho\otimes \rho$ is sufficient to accurately estimate $C_\rho(\alpha)C_\rho(-\bar{\alpha})$.   The details of the proof  and  its extension to $k>1$ modes are provided in the appendix.

%This idea of proof for single-mode state can be easily extended to multi-mode state by noting that $D(\bm{\alpha})=D(\alpha_1)\otimes \cdots\otimes D(\alpha_k)$ for $\bm{\alpha}\in \mathbb C^k$. %For each two copies $\rho\otimes \rho$, we apply $k$ balanced beam splitters pairwise on the $k$ pairs of modes and then perform homodyne measurements on each pair of outputs with one on position and the other on momentum.

For states with reflection symmetry, the estimation of the product $C_\rho({\bm{\alpha}})C_\rho(-\bar{\bm{\alpha}})$ is equivalent to the estimation of the square of the characteristic function $C_\rho({\bm{\alpha}})^2$. This fact is evident for states satisfying the condition   $C_\rho(-\bar{\bm{\alpha}})=C_\rho({\bm{\alpha}})$. 
The more general case of states satisfying the condition $C_\rho  (-\overline{\bm \alpha})   =  C_\rho  (  {\bm \alpha}   U)$ for some unitary matrix $U$   is treated in the appendix.  In turn,  the estimation of $C_\rho({\bm{\alpha}})^2$ can be used to estimate the purity $\tr(\rho^2)=1/\pi^{2k}\int_{\mathbb C^k} d^{2k}\bm{\alpha} |C_\rho(\bm{\alpha})^2|$. % {\color{blue}  [I am a bit confused here: when there is reflection symmetry, we can estimate $C_\rho(\bm{\alpha})^2$, but in the purity we have $|C_\rho(\bm{\alpha})|^2$. These two quantities are equal when $C_\rho(\bm{\alpha}$ is real, but what if it isn't?]     or other nonlinear functionals containing even powers of the characteristic function.} 

\begin{figure}
    \centering
    \includegraphics[width=0.48\textwidth]{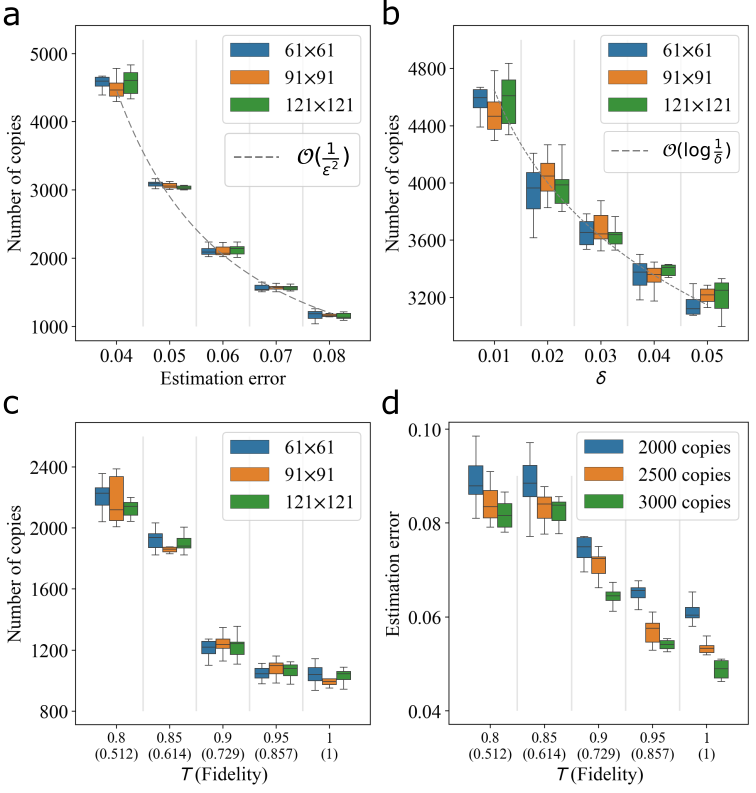}
    \caption{Estimation  of $C_\rho(\alpha)^2$ for the Fock state $\rho:=\ket{3}\bra{3}$. 
    Subfigure~a(b) shows how the required sample complexity scales with the estimation error (failure probability $\delta$) when $\delta=0.01$ (estimation error is $0.04$). 
    Subfigure~c and d show the effect of loss error on one copy of $\rho$ on the estimation results. Subfigure c shows how the required sample complexity scales with transmissivity rate $T$ (together with quantum fidelity between the noisy state and an ideal one) when $\epsilon=0.08$ and $\delta=0.02$, and Subfigure d shows how the estimation errors changes with $T$ when $\delta=0.02$.}
    \label{fig:fock}
\end{figure}

 A numerical experiment of estimation of $C_\rho({\alpha})^2$   for the Fock state $\ket{3}\bra{3}$ is shown in Figure~\ref{fig:fock}. There, the estimation error corresponds to the minimum value of $\epsilon$ satisfying $|\widehat{C_\rho(\alpha)^2}-C_\rho(\alpha)^2|\le \epsilon$ for a fraction $1-\delta$ of the phase space points on a grid. We consider grids of three different sizes: $61\times 61$, $91\times 91$ and $121\times 121$ within the region $[-3, 3]\times [-3, 3]$ in phase space, where each intersection corresponds one phase space point under consideration.
  Notice that, up to statistical fluctuations, the estimation error is independent of number of phase space points, in agreement with  Lemma \ref{lemma1}.  The figure also shows the robustness of our method to deviations from the condition of perfect copies.  Specifically,  we considered the scenario in which one of the two copies undergoes photon loss,  and observed that the estimation appears to be robust to small losses (Subfig.~\ref{fig:fock}c).

\begin{figure}
    \centering
    \includegraphics[width=0.48\textwidth]{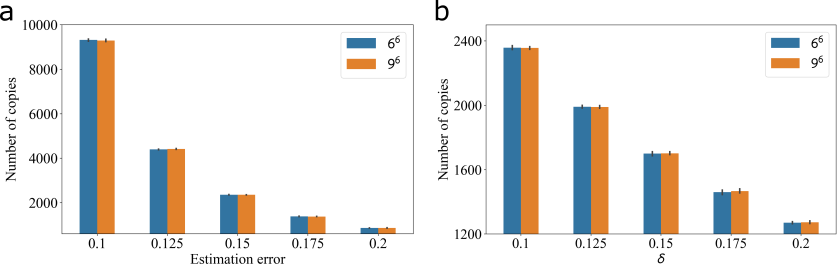}
    \caption{Box plots regarding estimation of $C_\rho(\bm{\alpha})$ for a three-mode  one-dimensional CV cluster state with squezzing parameter $0.9$. Here the estimation error is defined as the minimum value of $\epsilon$ satisfying $|\widehat{C_\rho(\bm{\alpha})}-C_\rho(\bm{\alpha})|\le \epsilon$ for $(1-\delta)\times 100\%$  of phase space points $\bm{\alpha}$ on a grid in the region $[-3, 3]^6$ of the six-dimensional phase space. Figure a shows how the required sample complexity scales with the estimation error when $\delta=0.1$ and Figure b shows how the sample complexity scales with the failure probability $\delta$ when $\epsilon=0.15$ for two grids of  sizes $6^6=46656$ and $9^6=531441$. }
    \label{fig:3mode}
\end{figure}

Lemma \ref{lemma1} has an  important implication:  if we know  that the characteristic function of the state is  has reflection symmetry,  and, in addition, is positive, then we can estimate its value at $M$ phase points with $O(\log M)$ state copies in an experimentally feasible approach.
\begin{cor}\label{cor:positive}
    For every $k$-mode state $\rho$ with  reflection symmetry and positive characteristic function,  the values  of the characteristic function   at  $M$ given points  can be estimated from  $O(\log(M/\delta)/\epsilon^4)$ copies using only beamsplitters and homodyne measurements.
    %so the probability that the error is larger then $\epsilon$ is upper bounded by $1-\delta$. 
\end{cor}
This result can be used to estimate the characteristic function of squeezed vacuum states with known phase, both in the single-mode and in the multimode scenario.   A numerical experimement of estimation of the characteristic function of a three-mode CV one-dimensional cluster state is provided in  Figure~\ref{fig:3mode}.  

Let us consider now the general case where the characteristic function can take arbitrary complex values. In this case, the square $C_\rho({\bm{\alpha}})^2$ determines the value of the characteristic function up to a sign.  The second technique used in the derivation of Theorem \ref{thm} is a method for identifying the correct sign of the characteristic function.

\begin{lemma}\label{lemma2}
  Let $\rho$ be  a   $k$-mode CV state  and let  $\{\bm{\alpha}_i\}_{i=1}^L$ be a set of phase space points satisfying the condition  $|C_\rho(\bm{\alpha_i})|>\epsilon$ for every  $i\in  \{1,\dots, L\}$.    Then,  the signs of all $C_\rho(\bm{\alpha_i})$ can be estimated from  $O(1/\epsilon^2\log (L/\delta))$ copies of $\rho$ with probability of error at most $\delta$.  
\end{lemma}

Lemma~\ref{lemma2} utilizes joint measurements performed on many copies of a CV state (Figure \ref{fig:my_label}b), the implementation of which may require scalable universal quantum computers.
Combining Lemma~\ref{lemma1} and Lemma~\ref{lemma2}, we then obtain Theorem \ref{thm}. To estimate the characteristic function at each of $M$ phase space points up to error $\epsilon$, we first estimate its square $C_\rho(\bm{\alpha})^2$ up to error $O(\epsilon^2)$, using the technique provided by Lemma~\ref{lemma1}.   This step requires  $O(  \log(M/ \delta)/\epsilon^4)$ copies of the state $\rho$.  We then check whether the modulus of the estimate is close to zero for the $M$ values of interest.  If $\left|\widehat{C_\rho({\bm{\alpha}_i})^2}\right|$ is  less than $4\epsilon^2/9$, we set the estimate  of the characteristic function to zero, namely  $\widehat{C_\rho({\bm{\alpha}_i})}=0$. Otherwise, we can estimate the sign of the characteristic function.  By  Lemma~\ref{lemma2},  this  step consumes $O(\epsilon^{-2}  \log(M/\delta)  )$.     The full details of the proof are provided in the appendix.

 Now let us discuss the  computational complexity of our approach.
For each phase space point $\bm \alpha$, the value of  $C_\rho(\bm \alpha)^2$ is estimated from the empirical average in Eq.~(\ref{eq:empiricalAvg}), which can be computed in  $O(\log M)$  time.  On the other hand, the computational time for estimating the sign of $C_\rho(\bm \alpha)$ is a constant. Hence, the computational complexity for estimating all the values of $C_\rho(\bm \alpha)$ for $M$ phase-space points is $O(M\log M)$.

%The estimation errors in all cases are bounded above by $\epsilon$ with probability at least $1-\delta$. The overall sample complexity is bounded by $N_1+N_2=O(1/\epsilon^4\log(M/\delta))$.

\medskip

{\em Comparison with other approaches. } Theorem \ref{thm} shows that the characteristic function at $M$  points can be accurately estimated using global measurements on $O( \log M)$ copies of the state, as illustrated in Figure \ref{fig:my_label}b.   This setting is different from that of conventional scenarios in which each copy of the state undergoes an individual measurement (Figure \ref{fig:my_label}c). Consider for example the naive conventional scenario in which each copy is used to estimate  the value of the characteristic function value at one  specific  point.  Intuitively,   estimating  $M$ different values in this naive setting will require a number of samples growing linearly in $M$, no matter what kind of classical post-processing is done on the experimental data.    This intuition can be made rigorous using  the results of Ref.~\cite{huang2021} on the complexity of learning point functions.     This result can be summarized in the following proposition.
\begin{proposition}\label{prop}
For a reflection symmetric CV state, the sample complexity of the estimation of the characteristic function  at $M$  points   up to a constant error with high probability  is at least $\Omega(M)$  using individual measurements in the naive  scenario.
\end{proposition}

Our method also exhibits   advantages   over classical shadow tomography. When used for estimating a set of observables over a $k$-mode CV state, existing methods of classical shadow tomography using homodyne measurements~\cite{gandhari2024,becker2024} have a sample complexity growing exponentially with $k$, in contrast with the sample complexity of our method, which is independent of $k$.      Moreover, classical shadow tomography approaches require a truncation, either in Fock space  or the phase space, which is not necessary in our method for estimation of point values of a state characteristic function. % {\color{blue}  By  avoiding the truncation, our method significantly reduces the scaling of the computational complexity with $M$, bringing it down from [???] to $O(M \log M)$.}

\section{Application: estimation of  CV observables}   
Our method for estimating the characteristic function can be used to estimate the expectation value of a variety of CV observables.   In general, the expectation value of a $k$-mode observable $O$ on a state $\rho$ is given by the tomographic formula $\tr [  O  \, \rho]  =  \int \textrm{d}^{2k}\bm{\alpha}  \,  C_\rho(\bm{\alpha})  \,  C_O(-\bm{\alpha})/\pi^k$, with $C_O(\bm \alpha)  :  = \tr[   O \, D  (  -  i \bm \alpha)]$~\cite{serafini2017}.   Now, suppose that the observable has bounded trace norm $|| O||_1\le 1$ and satisfies the condition    $|\int_{\bm{\alpha}\notin \mathcal A} \textrm{d}^{2k}\bm{\alpha} \, C_\rho(\bm{\alpha}) \,C_O(-\bm{\alpha})|<\epsilon/2 $ for some  $\epsilon>0$ and  some compact region $\mathcal A\subset  \mathbb C^k$. For example, this  condition is satisfied if $C_O(\bm \alpha)$ decays exponentially with $|\bm \alpha|$, as it happens {\em e.g.} when $O$ is the density operator of a  $k$-mode coherent quantum state  and the region $\mathcal A$ is large compared to the amplitude of such state.  In this case, an estimate of the expectation value of $O$ can be obtained by  randomly sampling $M$ points inside $\mathcal A$ and by estimating the characteristic function of $\rho$ at these points. In the appendix, we show that picking    $M=16\sigma_M^2 |\mathcal A|^2/\epsilon^2$, where $\sigma_M^2:=\frac{1}{M-1}\sum_{i=1}^M \left( \widehat{C_\rho(\bm \alpha_i)}C_O(-\bm \alpha_i)- \frac{1}{M}\sum_{i=1}^M \widehat{C_\rho(\bm \alpha_i)}C_O(-\bm \alpha_i) \right)^2$, $|\mathcal A|$ is the volume of $\mathcal A$, and estimating the characteristic function with error $\tilde{\epsilon}=\epsilon/(4|\mathcal A|)$ guarantees an accurate estimate of $\tr (O \rho)$.  
\begin{cor}
      The expectation value of a $k$-mode observable $O$ with $||O||_1\le 1$ on a state $\rho$ can be estimated  using $O\left(|\mathcal A|^4/\epsilon^4\log (|\mathcal A|^2 /(\epsilon^{2}\delta))\right)$ copies of $\rho$ and the estimate $o:=\frac{|\mathcal A|}{\pi^2 M}\sum_{i=1}^M \widehat{C_\rho(\bm{\alpha}_i)} C_O(\bm{\alpha}_i)$  satisfies the condition 
    ${\sf Prob}\left(\left|o-\tr(\rho O)\right|\gtrsim\epsilon\right)< \delta$, where  $\sim$ comes from the approximation of estimation error of Monte Carlo integration.
\end{cor}
Note also that the same randomly sampled points can be used for any number of observables $O$, provided that all observables have a small contribution outside the region $\mathcal A$. This means that we don't need to resample and use more $\rho$ states, and the sample complexity only depends on the size of the region $|\mathcal{A}|$. % Hence, the sample complexity of the estimation depends on the region $\mathcal A$, and is independent of the number of observables.

\medskip

{\em Discussion and conclusions.}
Recently, many efforts have been made to efficiently learn continuous-variable quantum systems~\cite{iosue2024,gandhari2024,becker2024,oh2024,mele2024}. In this context, we have demonstrated that the characteristic function of a  multimode state with reflection symmetry can be  estimated at $M$ points using  $O(\log M)$ copies of  the state, independently of the number of modes.  
This contrasts with the naive conventional scenario, where $\Omega(M)$ copies of $\rho$ are required. For states with positive characteristic function, such as squeezed vacuum states,  the estimation is achieved by an experimentally-friendly setup that uses only beamsplitters and homodyne measurements.    Our approach does not require truncation of the Hilbert space and the sample complexity is independent of the number of modes, thereby enabling applications to  the  characterization of large-scale CV quantum states. Finally, regarding the fact that Gaussian states~\cite{bartlett2002} and particular types of nonGaussian states~\cite{calcluth2022} can be efficiently simulated on a classical computer, it is an interesting open problem whether the CV quantum states with reflection symmetry considered in this paper are efficiently simulatable. 

%Furthermore, for any CV state $\rho$ having certain known reflection symmetry, we only need a constant number of copies of $\rho$ to estimate $C_\rho(\bm{\alpha})^2$ for arbitrary phase-space point $\bm{\alpha}\in\mathbb C^k$ with experimentally friendly measurements.
%Based on this result, we have proposed a feasible proposal using squeezed-vacuum states, beam splitters and homodyne measurements, for experimentally demonstrating the quantum advantage in learning CV states.
%Finally, it is open whether the exponential quantum advantage still exists against general conventional approaches including adaptive strategies, where we can change the measurement setting on a next quantum copy based on previous measurement outcomes, for obtaining the state characteristic function point values. We also leave this question for future study.

\section{Acknowledgement}
This work was supported by funding from the Hong Kong Research Grant Council through grants no.\ 17300918 and no.\ 17307520, through the Senior Research Fellowship Scheme SRFS2021-7S02, the Croucher Foundation, and by  the John Templeton Foundation through the ID No. 62312 Grant, as part of the “The Quantum Information Structure of Spacetime” Project (QISS).
 Research at the Perimeter Institute is supported by the Government of Canada through the Department of Innovation, Science and Economic Development Canada and by the Province of Ontario through the Ministry of Research, Innovation and Science. This research is also supported by the Science and Technology Program of Shanghai, China (21JC1402900), NSFC grant No. 12341104, the Shanghai Jiao Tong University 2030 Initiative, and the Fundamental Research Funds for the Central Universities.  The opinions expressed in this publication are those of the authors and do not necessarily reflect the views of the John Templeton Foundation.

\bibliography{refs}

\appendix
\section{Preliminary knowledge of CV states and characteristic functions}
To make our paper complete, we present some useful preliminary knowledge about CV quantum states and characteristic functions here. However, for comprehensive understanding of CV quantum optics in phase space, we refer the readers to Refs.~\cite{scully1997,leonhardt1997,serafini2017}.
For the definition of characteristic functions, we follow the convention in Ref.~\cite{leonhardt1997}.
Let us first define displacement operators
\begin{equation}
    D(\alpha):=\textrm{e}^{\alpha\hat{a}^\dagger -\bar{\alpha}\hat{a}}, 
\end{equation}
where $\alpha\in\mathcal C$, $\hat{a}$ and $\hat{a}^\dagger$ are the annihilation operator and creation operator respectively.
It is a unitary operation $D(\alpha)^{-1}=D(\alpha)^\dagger=D(-\alpha)$.
%$D(\alpha)$ is named displacement operator as it yields a shift on the phase space with an amount of $\alpha$ due to the fact that $D(\alpha)\hat{a}D(-\alpha)=\hat{a}-\alpha$.
Two displacement operators do not commute with each other in general, satisfying 
\begin{align*}
    D(\alpha)D(\beta)=&\textrm{e}^{[\alpha\hat{a}^\dagger-\bar{\alpha}\hat{a}, \beta\hat{a}^\dagger-\bar{\beta}\hat{a}]} D(\beta)D(\alpha)\\
    =&\textrm{e}^{\alpha\bar{\beta}-\bar{\alpha}\beta} D(\beta)D(\alpha),
\end{align*}
since $[\hat{a}, \hat{a}^\dagger]=1$.
The set of all displacement operators form a Heisenberg-Weyl group~\cite{bartlett2002}, which is the generalization of Pauli group in CV quantum systems.

For any Hermitian trace-class operator $O$ on the Hilbert space $\mathcal H\cong L^2(\mathbb R)$, we have the Fourier-Weyl relation
\begin{equation}
    O=\frac{1}{\pi}\int_{\mathbb C} \textrm{d}^2 \alpha \tr(O D(\alpha)) D(-\alpha).
\end{equation}
For a density operator $\rho$ on $\mathcal H$, we immediately have
\begin{equation}
    \rho=\frac{1}{\pi}\int_{\mathbb C} \textrm{d}^2 \alpha \tr(\rho D(\alpha)) D(-\alpha).
\end{equation}
We define the state characteristic function as
\begin{equation}\label{eq:defCharacteristic}
    C_\rho(\alpha):=\tr(\rho D(-\textrm{i}\alpha))=\tr(\rho\textrm{e}^{-\textrm{i}\alpha\hat{a}^\dagger-\textrm{i}\bar{\alpha}\hat{a}}).
\end{equation}
(In some literature, the characteristic function of $\rho$ is defined as $\tr(\rho D(\alpha))$. Here we choose the convention in (\ref{eq:defCharacteristic}) due to the convenience when we consider measuring $C_\rho(\alpha)$.) 
In the following, sometimes we use $C_\rho(x, p)$ instead of $C_\rho(\alpha)$ to denote the characteristic function on the two-dimensional coordinate system by changing the variable $\alpha=\frac{x+\textrm{i}p}{\sqrt{2}}$.
In a similar way, we can define the observable characteristic function as
\begin{equation}
C_O(\alpha):=\tr(O D(-\textrm{i}\alpha)).
\end{equation}
Since displacement operators form an orthogonal basis on $\mathcal H$, we have
\begin{equation}
    \tr(\rho O)=\frac{1}{\pi^2} \int \textrm{d}^2 \alpha C_\rho(\alpha) C_O(-\alpha).
\end{equation}

It is easy to find the characteristic function of any CV state $\rho$ should satisfy the properties
$C_\rho(0)=\tr(\rho)=1$ and $C_\rho(-\alpha)=\tr(\rho D(\textrm{i}\alpha))=\overline{\tr(\rho D(-\textrm{i}\alpha))}=\overline{C_\rho(\alpha)}$, where the bar denotes complex conjugate. Another property a characteristic function must satisfy due to the semi-definite positivity of $\rho$ is given by quantum Bochner’s theorem~\cite{cushen1971}, which we will skip here.

Another important phase-space description of CV state $\rho$ is its Wigner quasi-probability function 
\begin{equation}
    W_\rho(x, p)= \frac{1}{2\pi}\int_{\mathbb R} \textrm{e}^{\textrm{i}p v}\left\langle x-\frac{v}{2}\left|\rho\right|x+\frac{v}{2}\right\rangle dv,
\end{equation}
for $x, p\in\mathbb R$.
Wigner function has an important operational property that is its marginal distribution equals the outcome probability distribution of a projected measurement on the associated quadrature basis, i.e.,
\begin{equation}\label{eq:relationMeasurementWigner}
    \textrm{P}_{\theta}(x)=\int_{\mathbb R} W_\rho(x \cos\theta- p\sin \theta, x\sin\theta+ p\cos \theta) d p,
\end{equation}
where $\textrm{P}_\theta$ is the outcome probability distribution of the homodyne measurement on quadrature $\cos\theta\hat{x}+\sin\theta \hat{p}$.
The characteristic function of a state $\rho$ can be obtained from its Wigner function by the Fourier transformation
\begin{equation}\label{eq:relationCharacteristicWigner}
    C_\rho(x, p)=\int_{\mathbb R^2} \textrm{d}u \textrm{d} v \textrm{e}^{-\textrm{i}(u x +v p)} W_\rho(u, v).
\end{equation}
Then for $\zeta\in\mathbb R$, we obtain
\begin{align*}
    &\int d x \textrm{P}_\theta (x) \exp(-\textrm{i}\zeta x)\\
    =& \int d x d p W_\rho(x \cos\theta- p\sin \theta, x\sin\theta+ p\cos \theta) \textrm{e}^{-\textrm{i}\zeta x} \\
    =&\int d x' d p' W_\rho(x', p') \textrm{e}^{-\textrm{i}\zeta(\cos\theta x'+\sin\theta p')}\\
    =&C_\rho(\zeta\cos\theta, \zeta\sin\theta)\\
    =&C_\rho(\zeta \textrm{e}^{\textrm{i}\theta}/\sqrt{2}),
\end{align*}
where we have used Eq.~(\ref{eq:relationMeasurementWigner}) in the first equality,  transformations of variables $x'=\cos\theta x-\sin\theta p$ and $p'=\sin\theta x+\cos\theta p$ in the second equality, and Eq.~(\ref{eq:relationCharacteristicWigner}) in the third equality.
By choosing $\theta=\textrm{arg}(\alpha)$ and $\zeta=\sqrt{2}|\alpha|$, we have 
\begin{equation}
    \int d x \textrm{P}_{\textrm{arg}(\alpha)} (x) \exp(-\textrm{i}\sqrt{2}|\alpha| x)=  C_\rho(\alpha).
\end{equation}
It implies that $C_\rho(\alpha)$ can be directly estimated by performing a homodyne detection in the quadrature basis with phase $\textrm{arg}(\alpha)$.

\section{Proof of Lemma 2}

Since balanced beam splitter yields the transformation
\begin{equation}
    \textrm{BS}\begin{bmatrix} \hat{a}_1 \\
    \hat{a}_2 
    \end{bmatrix}\textrm{BS}^\dagger=\begin{bmatrix}
    \frac{1}{\sqrt{2}}(\hat{a}_1+\hat{a}_2) \\
    \frac{1}{\sqrt{2}}(-\hat{a}_1+\hat{a}_2)
    \end{bmatrix},
\end{equation}
 we have
\begin{align*}\notag
    &\tr\left(D(-\textrm{i}\alpha)\otimes D(\textrm{i}\bar{\alpha})\rho\otimes\rho\right)\\\notag
    =&\tr\left( \textrm{BS} D(-\textrm{i}\alpha)\otimes D(\textrm{i}\bar{\alpha}) \textrm{BS}^\dagger \textrm{BS} \rho\otimes \rho \textrm{BS}^\dagger \right)\\
    =&\tr\left(D(-\textrm{i}\sqrt{2}\textrm{Re}(\alpha))\otimes D(\sqrt{2}\textrm{Im}(\alpha)) \textrm{BS} \rho\otimes\rho \textrm{BS}^\dagger \right).
\end{align*}
Note that the expectation value $\tr(D(-\textrm{i}\alpha/\sqrt{2})\rho)$ can be estimated by performing a homodyne detection in the quadrature basis with phase $\textrm{arg}(\alpha)$.  That is
\begin{equation}\label{eq:homodyneMeasureCharacteristic}
    \tr(D(-\textrm{i}\alpha/\sqrt{2})\rho)=\mathds{E}_x \exp(-i|\alpha| x),
\end{equation}
where $x$ denotes the homodyne measurement outcome.
\iffalse
This is because Wigner function $W_\rho$ is the Fourier transformation of the characteristic function $C_\rho$, i.e.
\begin{align}\notag
   C_\rho(\alpha)=&\tr(D(-\textrm{i}\alpha)\rho)\\ 
   =&\int_{\mathbb R^2} d x dp W_\rho(x,p) \exp(-\textrm{i}\sqrt{2}(x \textrm{Re}\alpha +p \textrm{Im}\alpha )),
\end{align}
where we have used the two-dimensional coordinate for $W_\rho$.
At the same time, the Wigner function is related to the probability distribution of the homodyne measurement outcome via~\cite{leonhardt1997}
\begin{equation}
    \textrm{P}_{\theta}(x)=\int W_\rho(x \cos\theta- p\sin \theta, x\sin\theta+ p\cos \theta) d p
\end{equation}
where $\textrm{P}_\theta (x)$ is the probability distribution of outcome $x$ of a homodyne measurement on quadrature with phase $\theta$ on state $\rho$.
\fi

To estimate $\braket{D(-\textrm{i}\sqrt{2}\textrm{Re}(\alpha))\otimes D(\sqrt{2}\textrm{Im}(\alpha))}_{\textrm{BS} \rho\otimes\rho \textrm{BS}^\dagger}$, we apply a balanced beam splitter to combine two copies $\rho\otimes \rho$, and then measure position at one mode  and measure momentum at the other mode.
When Re$(\alpha)$ is negative, we must flip the sign of the position measurement outcome when we plug the measurement outcome $x$ into the estimator $\mathbb E_{x} \exp(-\textrm{i}\sqrt{2}|\textrm{Re}(\alpha)|x)$ of the expectation value $\braket{D(\textrm{i}\sqrt{2}\textrm{Re}(\alpha))}$, but we do not need to flip the sign when Re($\alpha$) is positive. Similarly, when Im$(\alpha)$ is negative, we also need to flip the sign of momentum measurement outcome $p$ when we use the estimator $\mathbb E_{x} \exp(-\textrm{i}\sqrt{2}|\textrm{Im}(\alpha)|p)$, but we do not flip the sign when Im$(\alpha)$ is positive.
By denoting the measurement outcomes as $x$ and $p$ respectively (without flipping the signs), we have
\begin{align*}
    &\tr\left(D(-\text{i}\sqrt{2}\textrm{Re}(\alpha))\otimes D(\sqrt{2}\textrm{Im}(\alpha)) \textrm{BS} \rho\otimes\rho \textrm{BS}^\dagger \right)\\
    =&\mathds{E}_{x,p}\left(\exp(-2\textrm{i}\textrm{Re}(\alpha)x -2\textrm{i}\textrm{Im}(\alpha)p)\right).
\end{align*}
Suppose the total number of repetitions is $N$, then $\widehat{C_\rho(\alpha)C_\rho(-\bar{\alpha})}=\frac{1}{N}\sum_{i=1}^N \exp(-2\textrm{i}\textrm{Re}(\alpha)x^{(i)} -2\textrm{i}\textrm{Im}(\alpha)p^{(i)})$ is an unbiased estimator of $C_\rho(\alpha)C_\rho(-\bar{\alpha})$, where $x^{(i)}$ and $p^{(i)}$ are the $i$th position measurement outcome and momentum measurement outcome respectively.

Using Hoeffding's inequality for both the real and imaginary parts of the estimator, since both $|\textrm{Re}(\exp(-2\textrm{i}\textrm{Re}(\alpha)x^{(i)} -2\textrm{i}\textrm{Im}(\alpha)p^{(i)}))|\le 1$ and $|\textrm{Im}(\exp(-2\textrm{i}\textrm{Re}(\alpha)x^{(i)} -2\textrm{i}\textrm{Im}(\alpha)p^{(i)}))|\le 1$, we know that to make sure estimation error 
\begin{align*}
&| \widehat{C_\rho({\alpha})C_\rho(-\bar{\alpha})}- C_\rho(\alpha) C_\rho(-\bar{\alpha})|\\
\le &|\textrm{Re}(\widehat{C_\rho({\alpha})C_\rho(-\bar{\alpha})}- C_\rho(\alpha) C_\rho(-\bar{\alpha}))| \\
&+ |\textrm{Im}(\widehat{C_\rho({\alpha})C_\rho(-\bar{\alpha})}- C_\rho(\alpha) C_\rho(-\bar{\alpha}))|\\
\le& \epsilon
\end{align*}
with probability $1-\delta$, $N=O(\log(1/\delta)/\epsilon^2)$ is sufficient. 

For $k$-mode state, a displacement operator on the $2k$-dimensional phase space is a tensor product of single-mode displacement operators, i.e,
$D(\bm{\alpha})=D(\alpha_1)\otimes \cdots \otimes D(\alpha_k)$ for any $\bm{\alpha}=(\alpha_1, \dots, \alpha_k)\in \mathbb C^k$. Then following the above analysis, we have
\begin{align*}
&C_\rho(\bm{\alpha})C_\rho(-\bar{\bm{\alpha}})\\
=&    \braket{\otimes_{i=1}^k \left(D(-\textrm{i}\sqrt{2}\textrm{Re}(\alpha_i))\otimes D(\sqrt{2}\textrm{Im}(\alpha_i))\right)}_{\textrm{BS}^{\otimes k} \rho\otimes \rho \textrm{BS}^{\dagger \otimes k} },
\end{align*}
where each beam splitter operation is applied pairwise on the same modes of two copies $\rho\otimes \rho$. By performing homodyne measurements on each of the $2k$ modes obtaining outcomes $\{x_i\}_{i=1}^k$ and $\{p_i\}_{i=1}^k$, we can estimate $C_\rho(\bm{\alpha})C_\rho(-\bar{\bm{\alpha}})$ by 
\begin{align*}
    &\mathbb{E}_{x_1, \dots, x_k, p_1, \dots, p_k} \Pi_{i=1}^k \textrm{e}^{-2\textrm{i}\textrm{Re}(\alpha_i)x_i} \textrm{e}^{-2\textrm{i}\textrm{Im}(\alpha_i)p_i}\\
    =&\mathbb{E}_{x_1, \dots, x_k, p_1, \dots, p_k} \exp\left(\sum_{i=1}^k -2\textrm{i} (\textrm{Re}(\alpha_i)x_i+\textrm{Im}(\alpha_i)p_i)\right).
\end{align*}

\section{Proof of Corollary 1}

There is a unitary operation $\mathcal U$ on Hilbert space $\mathcal H$ that yields the transformation $\mathcal U\hat{\bm{a}}\mathcal U^\dagger= U\hat{\bm{a}}$ and $\mathcal U\hat{\bm{a}}^\dagger\mathcal U^\dagger= U^\dagger \hat{\bm{a}}^\dagger$, where $\mathcal U\hat{\bm{a}} \mathcal U^\dagger$ denotes the application of $\mathcal U\cdot \mathcal U^\dagger$ entrywise on vector $\hat{\bm{a}}$, and $U\in \mathbb C^{k\times k}$ is a unitary matrix. Using the above fact, we have 
\begin{align*}
\mathcal U D(\text{i}\bar{\bm{\alpha}}U) \mathcal U^\dagger =&\textrm{e}^{ \text{i}\bar{\bm{\alpha}}U \mathcal U\hat{\bm{a}}^\dagger\mathcal U^\dagger-\text{i} \bm{\alpha} U^\dagger \mathcal U \hat{\bm{a}} \mathcal U^\dagger}\\
=& \textrm{e}^{\text{i}\bar{\bm{\alpha}}\hat{\bm{a}}^\dagger-\text{i}\bm{\alpha}\hat{\bm{\alpha}}}\\
=&D(\text{i}\bar{\bm{\alpha}}).
\end{align*}
Hence
\begin{align*}
    C_{\rho }(-\bar{\bm{\alpha}}U)=&\tr(\mathcal U \rho \mathcal U^\dagger \mathcal U D(\textrm{i}\bar{\bm{\alpha}}U)\mathcal U^\dagger)\\
    =&\tr(\mathcal U \rho \mathcal U^\dagger D(\textrm{i}\bar{\bm{\alpha}})) \\
    =&C_{\mathcal U\rho \mathcal U^\dagger}(-\bar{\bm{\alpha}}).
\end{align*}
By using the condition of reflection symmetry that is $C_\rho(\bm{\alpha})=C_\rho(-\bar{\bm{\alpha}}U)$, we obtain $C_{\rho }(\bm{\alpha})^2=C_\rho(\bm{\alpha})C_{\mathcal U\rho \mathcal U^\dagger}(\bar{\bm{\alpha}}) $.
It implies that given two copies $\rho\otimes\rho$, after we perform a unitary operation $\mathcal U$ consisting of beam splitters and phase shifters on the second copy, we can follow the previous procedure explained in Lemma~1 to obtain an accurate estimation of $C_\rho(\bm{\alpha})^2$ using $O(1/\epsilon^2\log 1/\delta)$ copies of $\rho$ for each arbitrary point $\bm\alpha$. To make the estimations at all $M$ phase space points accurate, the union bound indicates that we need $O(1/\epsilon^2\log M/\delta)$ copies.

\section{Proof of Lemma 3}

First we suppose $|\textrm{Re}(C_\rho(\bm{\alpha_i}))|>\epsilon$ for all $\bm{\alpha}_i\in\mathcal S$.
Given $N$ copies of CV state $\rho$, that is $\rho^{\otimes N}$, for each $\bm{\alpha}\in\mathbb C^k$, to determine the sign of Re$(C_\rho(\bm{\alpha}))$, we measure the following two-outcome observable 
\begin{equation}\label{eq:twoOutcomeMeasurement}
   \int d \bm{q} f(\bm{q})\ket{\bm{q}_1}\bra{\bm{q}_1}\otimes\ket{\bm{q}_2}\bra{\bm{q}_2}\otimes \dots\otimes \ket{\bm{q}_N}\bra{\bm{q}_N},
\end{equation}
where $\bm{q}=(\bm{q}_1, \dots, \bm{q}_N)$, each $\ket{\bm{q}_i}=\ket{q_{i_1}}\otimes \cdots \otimes \ket{q_{i_k}} (1\le i\le N)$ is a tensor product of the eigenstates of the quadrature operators with phases $\textrm{arg}(\alpha_1), \dots, \textrm{arg}(\alpha_k)$ with eigenvalue $q_{i_1}, \dots, q_{i_k}\in \mathbb R$, and 
\begin{equation}\label{eq:signmeasurement}
    f(\bm{q})=\begin{cases}
        1, & \frac{1}{N}\sum_{i=1}^N\cos\left(\sqrt{2}\sum_{j=1}^k|\alpha_j|q_{i_j}\right)>0 \\
        0, & \frac{1}{N}\sum_{i=1}^N \cos\left(\sqrt{2}\sum_{j=1}^k|\alpha_j|q_{i_j}\right)\le 0
    \end{cases}.
\end{equation}
Note that because
\begin{align*}
    &C_\rho(\bm{\alpha})\\
    =&\mathbb{E}_{q_1,\dots,q_k} \exp\left(\sum_{i=1}^k -\sqrt{2}\textrm{i} |\alpha_i|q_i\right)\\
    =&\mathbb{E}_{q_1,\dots,q_k} \left(\cos\left(\sqrt{2}\sum_{i=1}^k|\alpha_i|q_i\right)-\textrm{i}\sin\left(\sqrt{2}\sum_{i=1}^k|\alpha_i|q_i\right)\right),
\end{align*}
we have
\begin{equation}
    \textrm{Re}(C_\rho(\bm{\alpha}))=\mathbb{E}_{q_1,\dots,q_k} \cos\left(\sqrt{2}\sum_{i=1}^k|\alpha_i|q_i\right).
\end{equation}
It implies that the measurement in (\ref{eq:twoOutcomeMeasurement}) is an estimation of the sign of Re($C_\rho(\bm{\alpha})$).
When the measurement outcome is $1$, then we believe Re$(C_\rho(\bm{\alpha}))$ is positive; otherwise, we believe Re$(C_\rho(\bm{\alpha}))$ is negative.

Without loss of generality, we assume that Re$(C_\rho(\bm{\alpha}))$ is positive and then calculate the probability to falsely determine Re$(C_\rho(\bm{\alpha}))$ as negative.
Using Hoeffding's inequality, we know that
the probability to obtain outcome $0$ in (\ref{eq:signmeasurement})  is bounded by
\begin{align*}
     &\textrm{Pr}\left(\frac{1}{N}\sum_{i=1}^N\cos\left(\sum_{j=1}^k|\alpha_j|q_{i_j}\right)\le 0\right)\\
     \le & \exp\left(\frac{-N \cdot \textrm{Re}(C_\rho(\bm{\alpha}))^2}{2}\right)\\
    \le &\textrm{e}^{-\frac{N\epsilon^2}{2}}.
\end{align*}
Hence, the error probability for determining the sign of Re$(C_\rho(\bm{\alpha}))$ by performing the above joint measurement on $\rho^{\otimes N}$ is bounded above by $\textrm{e}^{-N\epsilon^2/2}$.

Measuring the observable in (\ref{eq:twoOutcomeMeasurement}) only yields the majority vote of the sign. When $N$ is sufficiently large, $\rho^{\otimes N}$, with a high probability, falls inside either one of two eigenspaces of the observable. It implies that measuring the observable in (\ref{eq:twoOutcomeMeasurement}) does not disturb $\rho^{\otimes N}$ with a high probability and suggests that we can sequentially perform such quantum measurements for different values of $\bm{\alpha}$ on $\rho^{\otimes N}$.
Next we use the quantum union bound~\cite{gao2015,o2022}:
for any quantum state $\rho$ and a set of $M$ two-outcome observables $\{K_i\}_{i=1}^M$, each of which has eigenvalue $0$ or $1$. If $\tr(K_i \rho) \ge 1 -\epsilon$, then when we measure $K_1, \dots, K_M$ sequentially on $\rho$, the probability that all of them yield the outcome $1$ is at least $1 - 4M\epsilon$.
The quantum union bound implies that if we perform $M$ two-outcome measurements shown in (\ref{eq:twoOutcomeMeasurement}) for $M$ phase-space points $\{\bm{\alpha}_i\}_{i=1}^M$ sequentially on $\rho^{\otimes N}$, the error probability to obtain all the signs of $\{\textrm{Re}(C_\rho(\bm{\alpha}_i))\}_{i=1}^M$ is upper bounded by $4M\textrm{e}^{-N\epsilon^2/2}$.
Hence, if we choose $N=O(1/\epsilon^2\log (M/\delta))$, then we can correctly determine all the signs of $\{\textrm{Re}(C_\rho(\bm{\alpha}_i))\}_{i=1}^M$ with error probability less than $\delta$.  

Second, we suppose $|\textrm{Im} (C_\rho(\bm{\alpha}_i))|\ge \epsilon$ for all $\bm{\alpha}_i\in\mathcal S$. For each $\bm{\alpha}$, to determine the sign of Im($\bm{\alpha}$), we perform the two-outcome measurement in (\ref{eq:twoOutcomeMeasurement}) on $\rho^{\otimes N}$ with $f(\bm{q})$ redefined as following
\begin{align}
    f(\bm{q})=\begin{cases}
        1 & \frac{1}{N}\sum_{i=1}^N\sin(\sum_{j=1}^k|\alpha_j|q_{i_j})<0 \\
        0 & \frac{1}{N}\sum_{i=1}^N \sin(\sum_{j=1}^k|\alpha_j|q_{i_j})\ge 0
    \end{cases}.
\end{align}
When the outcome is $1$, we believe Im$(C_\rho(\bm{\alpha}))$ is positive, and otherwise, we believe Im($C_\rho(\bm{\alpha})$) is negative. This is because $\textrm{Im}(C_\rho(\bm{\alpha}))=-\mathbb E_{q_1,\dots, q_k}(\sin(\sum_{i=1}^k|\alpha_i|q_i))$.
Similar to the above analysis, using Hoeffding's inequality, we know that the error probability to estimate the sign of Im$(C_\rho(\alpha))$ is bounded above by
\begin{equation}
     \exp\left(\frac{-N \cdot \textrm{Im}(C_\rho(\alpha))^2}{2}\right)
    \le \textrm{e}^{-\frac{N\epsilon^2}{2}}.
\end{equation}
Using the quantum union bound again, we know that if we perform $M$ such quantum measurements for $M$ values of $\bm{\alpha}$ on $\rho^{\otimes N}$ with $N=O(1/\epsilon^2\log (M/\delta))$, then we can correctly determine all the signs of $\{\textrm{Im}(C_\rho(\bm{\alpha}_i))\}_{i=1}^M$ with error probability less than $\delta$.

%When $|\textrm{Re}(C_\rho(\alpha))|\le\epsilon$ and $|Im (C_\rho(\alpha))|\le \epsilon$, we choose $C_\rho(\alpha)=0$.

\section{Proof of Theorem 1}

For any given set $\mathcal S$ of $M$ phase-space points, we first apply Lemma 1, together with the union bound, to obtain estimate 
$\widehat{C_\rho({\alpha})^2}:=|\widehat{C_\rho({\alpha})^2}|\textrm{e}^{\textrm{i}\theta}$ with $\theta\in [-\pi, \pi)$
of $C_\rho(\alpha)^2$ using $O(1/\epsilon^4\log (M/\delta))$ copies of $\rho$ such that
\begin{equation}\label{supeq:estimationsquare}
    Pr(\max_{\alpha\in\mathcal S}|\widehat{C_\rho({\alpha})^2}-C_\rho(\alpha)^2|>\epsilon^2/3)<\delta/2.
\end{equation} 
For each $\alpha\in \mathcal S$, if $|\widehat{C_\rho({\alpha})^2}|\le 4\epsilon^2/9$, then we choose our estimation of $C_\rho(\alpha)$ as $\widehat{C_\rho({\alpha})}=0$. Then with probability at least $1-\delta/2>1-\delta$, we have
\begin{align*}
    |\widehat{C_\rho({\alpha})}-C_\rho(\alpha)|=& \sqrt{|C_\rho(\alpha)^2|} \\
    \le& \sqrt{|\widehat{C_\rho({\alpha})^2}|+| \widehat{C_\rho({\alpha})^2}-C_\rho(\alpha)^2|} \\
    \le& \sqrt{\frac{4\epsilon^2}{9}+\frac{\epsilon^2}{3} }\\
    <& \epsilon,
\end{align*}
where the first inequality is because of triangle inequality and the second inequality comes from (\ref{supeq:estimationsquare}).

Second, we consider the case $|\widehat{C_\rho({\alpha})^2}|>4\epsilon^2/9$.
By Lemma 3, we can use at most $O(1/\epsilon^2\log (M/\delta))$ copies of $\rho$ to correctly determine the sign of $C_\rho(\alpha)$ for at most $M$ phase-space points $\alpha$ with probability at least $1-\delta/2$. 
Denote the square root of $\widehat{C_\rho({\alpha})^2}$ closer to $C_\rho(\alpha)$ as $\sqrt{\widehat{C_\rho({\alpha})^2}}$.
 Then $|\sqrt{\widehat{C_\rho({\alpha})^2}}+C_\rho(\alpha)|>|C_\rho({\alpha})|>2\epsilon/3$, hence with probability at least $1-\delta/2$, we have
 \begin{align*}
    &|\sqrt{\widehat{C_\rho({\alpha})^2}}-C_\rho(\alpha)| \\
    =&\frac{|\widehat{C_\rho({\alpha})^2}-C_\rho(\alpha)^2|}{|\sqrt{\widehat{C_\rho({\alpha})^2}}+C_\rho(\alpha)|}\\
    \le &\frac{\epsilon^2/3}{ 2\epsilon/3}\\
    < &\epsilon.
\end{align*}
 Then using the triangle inequality, with probability $(1-\delta/2)^2>1-\delta$, we have
\begin{equation}
    |\widehat{C_\rho({\alpha})}-C_\rho(\alpha)| <\epsilon.
\end{equation}
Combining the above two cases, we have proved the main result.

\section{Example}
The implementation of joint measurements employed in Lemma 3 is beyond state-of-the-art quantum technology, and may require scalable universal quantum computers. However, in some particular cases, the measurement strategy for estimating the signs of $C_\rho(\alpha)$ can be significantly simplified. For instance, when the CV state under consideration is invariant under phase rotation, we can estimate the signs of $C_\rho(\alpha)$ for any $\alpha$ by using single one homodyne measurement setting and the sample complexity can be reduced to $O(\log(1/\delta) /\epsilon^2)$.

\begin{figure}[h]
    \centering
    \includegraphics[width=0.45\textwidth]{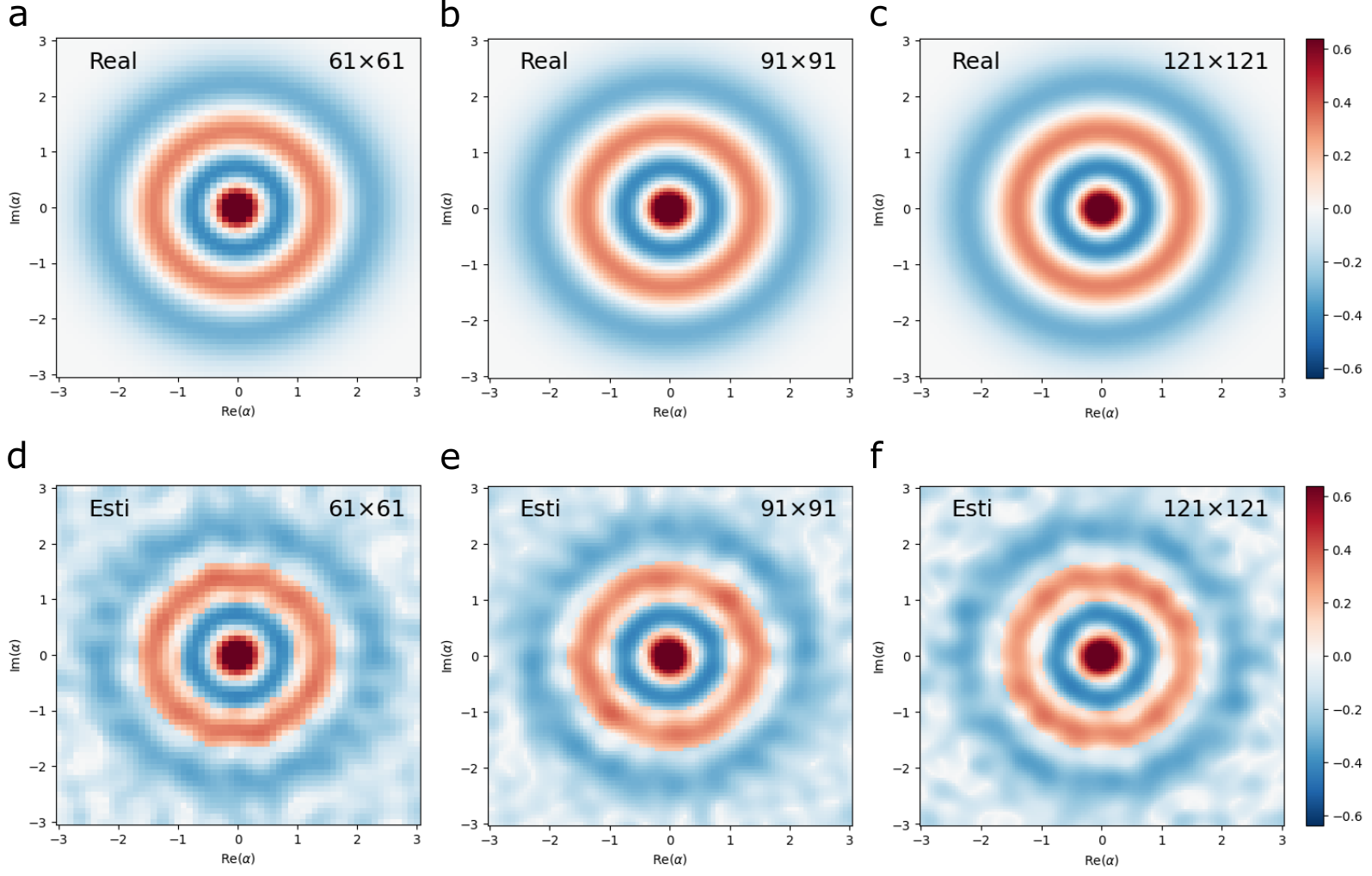}
    \caption{Comparison between true values of $C_\rho(\alpha)$ and the corresponding estimations $\widehat{C_\rho(\alpha)}$ for a Fock state $\rho=\ket{3}\bra{3}$. Figure a, b and c show the calculated true values of $C_\rho(\alpha)$ on grids of size $61\times 61$, $91\times 91$ and $121\times 121$ in the region $[-3, 3]\times [-3, 3]$ of phase space. Figure d, e, and f show the corresponding estimations $\widehat{C_\rho(\alpha)}$ on three grids respectively.}
    \label{fig:example}
\end{figure}

Here we consider a particular example of Fock state which is invariant under phase rotation.
 Figure~\ref{fig:example} shows the real parts of estimations of $C_\rho(\alpha)$, where $\rho$ is a Fock state $\rho=\ket{3}\bra{3}$ using $4100$ copies of $\rho$, together with the corresponding calculated truth. We use $4000$ copies of $\rho$ to obtain our estimation of $C_\rho(\alpha)^2$ for all $\alpha$. Specifically, since $C_\rho(\alpha)^2= C_\rho(\alpha)C_\rho(-\bar{\alpha})$ for a Fock state, we can use the setup illustrated by Fig.~1a to produce our estimation. We then apply a homodyne measurement on position over $100$ copies of $\rho$ to estimate the signs of $C_\rho(\alpha)$ for all $\alpha$. That is, for each $\alpha\in \mathbb C$, if $\text{Re}\left(\frac{1}{100}\sum_{i=1}^{100} \text{e}^{-\text{i}\sqrt{2}|\alpha|x_i}\right)>0$, where $x_i\in \mathbb R$ denotes a measurement outcome, then the estimate of $\textrm{sgn}(C_\rho(\alpha))$ is positive; otherwise, it is negative.

\section{Proof of Proposition 4}
Here we show that in restricted conventional scenario, at least $\Omega(M)$ copies of $\rho$ is required to estimate the values of characteristic function $\{C_\rho(\alpha_i)\}_{i=0}^{M-1}$ for $M$ phase-space points, even when state $\rho$ has knwon reflection symmetry. 
Since we consider the worst case, to prove this proposition, we only need to provide an example requiring sample complexity $\Omega(M)$.
Consider a set of single-mode squeezed-vacuum states $\{\rho_i\}_{i=0}^{2M-1}$, where each $\rho_i$ is squeezed in the quadrature  with phase $\theta_i=\frac{i\pi}{2M}$.
The characteristic function of each $\rho_i$ is $C_{\rho_i}(x, p)=\exp\left(-(x\cos\theta_i+p\sin\theta_i)^2/(4r^2)-r^2/4(x\sin\theta_i-p\cos\theta_i)^2\right)$.
Suppose  the given set of phase space points is $\mathcal S:=\{\alpha_i:=|\alpha|\textrm{e}^{\textrm{i} \theta_i}\}_{i=0}^{M-1}$ for certain fixed value of $|\alpha|> 0$. When squeezing asymptotically goes to infinity,
 we have $C_{\rho_i}(\alpha_i)\rightarrow
 1$ and for all $j\neq i$, $C_{\rho_i}(\alpha_j)\rightarrow 0$.

Now we define a new set of prepared states
$\{ \sigma_i |\sigma_i=\frac{1}{2} \rho_i+\frac{1}{2}\rho_{2M-i}\}_{i=0}^{M-1}$, each of which is a uniform mixture of two single-mode squeezed-vacuum states.
Due to the linearity of chracteristic functions, each $\sigma_i$ satisfies the reflection symmetry $C_\rho(\bm{\alpha})=C_\rho(-\bar{\bm{\alpha}})$ $\forall \bm{\alpha}$.  
When squeezing asymptotically goes to infinity, %such that $\textrm{e}^r\gg \max\left\{|\alpha|, 1/(|\alpha|\sin(\pi/2M))\right\}$, 
for any $0\le i,j\le M-1$, we have
\begin{equation}
    C_{\sigma_i}(\alpha_j)\begin{cases}
        > 1/3 & i=j \\
        <1/6 &  i\neq j.
    \end{cases}
\end{equation}
Suppose we are provided with copies of a state $\sigma_i$ randomly chosen from the set of $\{\sigma_i\}_{i=0}^{M-1}$.
Precisely estimating all the values of $\{C_{\sigma_i}(\alpha)\}$ for all $\alpha\in \mathcal S$ implies that we can correctly distinguish between $C_\rho(\alpha)> 1/3$ and $C_\rho(\alpha)<1/6$.

Then we can use the following lemma~\cite{huang2021} on the complexity of learning point functions to show that the sample complexity to accurately estimate the values of $\{C_{\sigma_i}(\alpha_j)\}_{j=0}^{M-1}$  in restricted conventional scenario is $\Omega(M)$.
\begin{lemma}\label{lemma:pointFunction}
    Suppose we have a set of functions $f_a(x)$ $(a=0,1,\dots, M-1)$ on $x_b$ ($b=0, 1, \dots, M-1$) satisfying 
    \begin{equation}
        f_a(x_b)=\begin{cases}
            1 & a=b \\
            0 & a\neq b.
        \end{cases}
    \end{equation}
    If there exists a classical randomized algorithm with input $\{x_{c_i}, f_a(x_{c_i})\}_{0\le c_i\le M-1, 1\le i\le N}$ and output $\tilde{f}_a$ such that $\forall 0\le a \le M-1$,
    \begin{equation}
        \textrm{Pr}\left(\max_{0\le b\le M-1}|f_a(x_b)-\tilde{f}_a(x_b)|>1/2\right)<1/3 ,
    \end{equation}
    then $N=\Omega(M)$.
\end{lemma}

From the above set of characteristic functions, we can easily obtain a set of point functions: when $C_{\sigma_i}(\alpha_j)>1/3 (<1/6)$, we set the point function $f_i(\alpha_j)=1(0)$. Hence, if we have a classical randomized algorithm using the measurement data in the restricted conventional scenario with sample complexity $o(M)$ to accurately estimate all the values of $\{C_{\sigma_i}(\alpha)\}$ for all $\alpha\in\mathcal S$, then we can correctly distinguish between the two cases $C_{\sigma_i}(\alpha)>1/3$ and $C_{\sigma_i}(\alpha)<1/6$ for each $\alpha\in\mathcal S$ with input size $o(M)$.
This of course contradicts with the above lemma. Thus, we know that the sample complexity to estimate values of $\{C_{\sigma_i}(\alpha)\}$ for all phase-space points $\alpha\in\mathcal S$ is at least $\Omega(M)$ in this restricted classical scenario.

\section{Proof of Corollary 2}
Suppose $k$-mode quantum state $\rho$ has finite energy in phase space such that there exists a region $\mathcal A\subseteq \mathbb C^k$ around the origin such that
$|\int_{\bm{\alpha}\notin \mathcal A} \textrm{d}^{2k} \bm{\alpha} C_\rho(\bm{\alpha})C_O(-\bm{\alpha})|<\epsilon/2 $.
Denote $\{\bm \alpha_i\}_{i=1}^M$ as the set of $M$ phase-space points we randomly sample in region $\mathcal A$ and $\widehat{C_\rho(\bm \alpha_i)}$ as the estimation of the state characteristic function value at phase-space point $\bm \alpha_i$ up to error $\tilde{\epsilon}$. 
Denote $\sigma_M^2:=\frac{1}{M-1}\sum_{i=1}^M \left( \widehat{C_\rho(\bm \alpha_i)}C_O(-\bm \alpha_i)- \frac{1}{M}\sum_{i=1}^M \widehat{C_\rho(\bm \alpha_i)}C_O(-\bm \alpha_i) \right)^2$ and $|\mathcal A|$ as the volume of region $\mathcal A$.
Then we have
\begin{align*}
    &\left|\int \textrm{d}^{2k}\bm \alpha C_\rho(\bm \alpha)C_O(-\bm \alpha)-\frac{|\mathcal A|}{M}\sum_{i=1}^M \widehat{C_\rho(\bm \alpha_i)}C_O(-\bm \alpha_i) \right|\\
    \le &\left|\int_{\bm \alpha\in\mathcal A} \textrm{d}^{2k}\bm \alpha C_\rho(\bm \alpha)C_O(-\bm \alpha)-\frac{|\mathcal A|}{M}\sum_{i=1}^M \widehat{C_\rho(\bm \alpha_i)}C_O(-\bm \alpha_i)\right|\\
    &+\left|\int_{\bm \alpha\notin \mathcal A} \textrm{d}^{2k}\bm \alpha C_\rho(\bm \alpha)C_O(-\bm \alpha)\right|\\
    \le &\left|\int_{\bm \alpha\in\mathcal A} \textrm{d}^{2k}\bm \alpha C_\rho(\bm \alpha)C_O(-\bm \alpha)-\frac{|\mathcal A|}{M}\sum_{i=1}^M C_\rho(\bm \alpha_i)C_O(-\bm \alpha_i)\right|\\
    & +\left|\frac{|\mathcal A|}{M}\sum_{i=1}^M (C_\rho(\bm \alpha_i)-\widehat{C_\rho(\bm \alpha_i)}) C_O(-\bm \alpha_i) \right| +\frac{\epsilon}{2}  
\end{align*}
where we have used the triangular inequality in the first and the second inequalities.

The first term is the estimation error in Monte Carlo integration, which can be approximated by $\frac{\sigma_M |\mathcal A| }{\sqrt{M}}$. 
To bound the second term, by using Holder's inequality and the fact that $\forall i$, $| C_\rho(\bm \alpha_i)-\widehat{C_\rho(\bm \alpha_i)}|\le \tilde{\epsilon}$ and $|C_O(-\bm \alpha_i)|\le ||D(\text{i}\bm \alpha_i)||_\infty ||O||_1\le ||O||_1\le  1$,
we have
\begin{align*}
& \left|\sum_{i=1}^M (C_\rho(\bm \alpha_i)-\widehat{C_\rho(\bm \alpha_i)}) C_O(-\bm \alpha_i) \right| \\
\le & \sum_{i=1}^M  |C_\rho(\bm \alpha_i)-\widehat{C_\rho(\bm \alpha_i)}| \max_{i=1,\dots, M} |C_O(-\bm \alpha_i)| \\
\le &M \tilde{\epsilon}
\end{align*}
Thus, we obtain
\begin{align*}
    &\left|\int \textrm{d}^{2k}\bm \alpha C_\rho(\bm \alpha)C_O(-\bm \alpha)-\frac{|\mathcal A|}{M}\sum_{i=1}^M \widehat{C_\rho(\bm \alpha_i)}C_O(-\bm \alpha_i) \right| \\
    \lesssim & \frac{\sigma_M|\mathcal A|}{\sqrt{M}}+|\mathcal A|\tilde{\epsilon}+\frac{\epsilon}{2}.
\end{align*}

To make the above estimation error less than $\epsilon$, we set $M=\frac{16\sigma_M^2|\mathcal A|^2}{\epsilon^2}$ and $\tilde{\epsilon}=\frac{\epsilon}{4|\mathcal A|}$. Inserting into the sample complexity in Theorem 3, we find that $O\left(\frac{|\mathcal A|^4}{\epsilon^4}\log\frac{\sigma_M^2 |\mathcal A|^2}{\epsilon^2\delta}\right)$ copies of $\rho$ are sufficient to estimate $\tr(\rho O)$ up to error $\epsilon$ with probability at least $1-\delta$.

\section{Discussion on preparing two copies of $\rho$}
Here we briefly discuss how to achieve two identical copies of a quantum state in experiments.
Ideally, an experimenter may control two individual quantum devices in a lab in the same way such that both these two devices produce the same quantum state. In practice, an experimenter can employ a quantum device to prepare the same quantum state simultaneously in two spatial modes. For example, one can prepare two spatial modes with the same two-mode squeezed state using four-wave mixing. 
It is also possible to consider that a quantum device keeps producing the same copy of a quantum state in each pulse. An experimenter can delay the transmission of one copy of the quantum state in one pulse through a waveguide and recombine two copies of the state in two different pulses later by a beam splitter. However, in this approach, the extra propagation of one pulse can introduce loss error in that corresponding copy of the quantum state. In the main text, we have considered how this kind of loss error can affect the efficiency of our approach as well as the resulting estimation errors. Numerical results imply that our approach is robust to low loss error occurred on one of two quantum copies.

\end{document}